\documentclass[12pt,a4paper]{article}
\usepackage[preprint]{dfttrob}
\usepackage{times}
\usepackage[numbers,sort&compress]{natbib}
\usepackage{latexuseful2e}
\usepackage{amsmath}
\usepackage{graphicx}

\newcommand{\bFB}{b_{FB}}   
\newcommand{\glog}{g_{\text{log}}}

\dfttnum{DFTT 34/2002}

\begin{document}

\title{Forward-Backward Multiplicity Correlations in Symmetric and
Asymmetric High Energy Collisions}

\author{Roberto Ugoccioni\\
\it Dipartimento di Fisica Teorica and I.N.F.N. - sezione di Torino \\
\it via P. Giuria 1, I-10125 Torino, Italy}

\maketitle

\begin{abstract}
Forward-backward correlations are explored within the
two-component clan model of multiparticle production. 
It is found that existing data are well described,
and, in $hh$ collisions, that clans must be allowed to leak
particles from one hemisphere to the other. General formulae given
for the symmetric case are then extended to the asymmetric one,
which is relevant for pA and AB collisions.
\end{abstract}


The analysis of multiplicity correlations between two different
regions of phase-space provides important information on 
the dynamics of multiparticle production processes. 

Due to space limitations, the reader is referred for all details
and references to papers \cite{RU:FB} and \cite{RU:FBasymm}.

I will at first consider symmetric reactions (like \ee\ annihilation,
$pp$ and $p\bar p$ collisions) with a symmetric choice of the
forward (F) and backward (B) regions: in \ee\ annihilations the F
regions is chosen randomly between those defined by a plane
through the collision point and perpendicular to the thrust axis;
in $hh$ collisions the collision axis is chosen as reference.
B is always the symmetric region.
The forward-backward (FB) correlation strength, $\bFB$, is defined as usual:
\begin{equation}
	\bFB =\frac{\avg{(n_F-\nbar_F)(n_B-\nbar_B)}}
			{\left[ \avg{(n_F-\nbar_F)^2} \avg{(n_B-\nbar_B)^2}
			\right]^{1/2}}  ,
\end{equation}
where $n_F$ if the number of particles in F and $n_B$ the number of
particles in B.
A more precise study of the correlations investigates the dependence
of the average number of particles in one hemisphere on the number of
particles in the opposite one, say $\bar n_F(n_B)$.

In $hh$ collisions the correlation strength grows 
logarithmically with the c.m.\ energy
$\sqrt{s}$ and is rather large (from 0.16 at 63 GeV to 0.7 at 1800
GeV); in \ee\ annihilation the
correlation strength is rather small (0.1 at 91 GeV).
Furthermore, $\bFB$ was measured in 2-jet and 3-jet events 
separately at LEP1 and was found to be compatible with zero in both cases.

The mechanism of the weighted superposition
of different classes of events, which was found to describe the
shoulder structure of multiplicity distributions and the oscillations
of their $H_q$ moments, can be tested on FB correlations.
The classes of events are defined according to
the reaction considered: in $hh$ collisions we define soft events
(events without mini-jets) as class 1,
and semi-hard events (events with mini-jets) as class 2;
in \ee\ annihilation,
class 1 are two-jet events, class 2 are three-jet events.

The starting point is the weighted superposition
equation for the joint distribution of $n_F$ and $n_B$
particles:
\begin{equation}
	P(n_F, n_B) = \alpha P_1(n_F, n_B) + (1-\alpha) P_2(n_F, n_B),
					\label{eq:combo}
\end{equation}
where $\alpha$ is the fraction of events of class 1.
Based on the above equation and symmetry properties,
it is possible to obtain explicitly 
a general formula for $\bFB$, which depends of course
on the first two moments of the general 
multiplicity distribution (MD) in each class 
$i=1,2$ (namely the
average multiplicity $\nbar_i$ and the dispersion $D_{n,i}$) and on the
correlation strength in each class, $b_i$:
\begin{equation}
	\bFB =  \frac{\alpha b_1 {D^2_{n,1}}(1+b_2) +
				(1-\alpha) b_2 {D^2_{n,2}}(1+b_1) +
					\frac{1}{2}\alpha(1-\alpha)(\nbar_{2} - \nbar_{1})^2(1+b_1)(1+b_2)}
			{\alpha  {D^2_{n,1}}(1+b_2) +
				(1-\alpha)  {D^2_{n,2}}(1+b_1) +
					\frac{1}{2}\alpha(1-\alpha)(\nbar_{2} -
				\nbar_{1})^2(1+b_1)(1+b_2)} .
				\label{eq:general}
\end{equation}

We apply first our result to OPAL data, setting $b_1 = b_2
= 0$ as the data suggest. We take the values of the moments of the MD
in each class from previous fits and obtain $\bFB = 0.101$ to be
compared with the OPAL result $0.103 \pm 0.007$.
Thus we conclude that the weighted superposition mechanism describes
well FB correlations in \ee\ annihilation.

We now turn to $p\bar p$ collisions at 546 GeV (UA5); the experimental
result is $0.58 \pm 0.01$. Taking the shoulder description from a fit
to the overall MD, one finds that Eq.~(\ref{eq:general})
with $b_1 = b_2 = 0$ gives a too small result, indicating
that FB correlations are needed in each substructure. On the other
hand, if particles in each substructure were totally uncorrelated,
one would get a result which is much larger than
experimentally found. We conclude that particles are emitted with
partial correlation, as already suggested by UA5.

Since we have succesfully used a two-step production process to
describe multiparticle dynamics, leading 
to the negative binomial
(NB) shape of the MD, and which naturally accommodates partial
correlation, we
apply it to the study of FB correlations in a generalised way:
\begin{itemize}
\item   In the first step, $N$ independent objects (generalised clans) are
  produced	according to a given MD, ${\cal P}(N)$.
\item   In the second step, particles are produced within each clan
  according to a MD $Q(n_c)$.
\end{itemize}
This two-step process is used to describe each component (with
possibly different parameters in each step): from now on, all our
formulae refer to each component separately, but we drop the component
index in the formulae.  We further assume that clans are produced
independently in rapidity thus the distribution in the number of
forward clans, $N_F$, at fixed total number of clan, $N$, is binomial.
The successful NB parametrisation of the data is obtained when
${\cal P}(N)$ is the Poisson distribution and $Q(n_c)$ is the
logarithmic distribution.

Consider now the situation in which  particles produced by an F (B)
clan all remain in the F (B) region: 
if clans follow a Poisson MD, as is implied by the very good fit
to the data obtained with the NB distribution (NBD), 
one 
obtains $b = 0$,
independently of the details of the MD inside clans. It is therefore
necessary to allow clans to leak particles from one hemisphere to the
other. In order to do this, a new parameter is introduced, $p$, which
is defined as the average fraction of particles in a clan which remain
in the same region and do not leak away to the opposite one. As clans
are classified F or B according to where the majority of the particles
fall, one has $0.5 \leq p \leq 1$. We also define $q = 1-p$.

The general result is now
\begin{equation}
	b = \frac{D^2_n/\nbar - D^2_c/\nc - (p-q)^2\nc + 4\gamma/\nc }{
				D^2_n/\nbar  + D^2_c/\nc +	(p-q)^2\nc - 4\gamma/\nc } ,
\end{equation}
where $\gamma$ is the covariance between F and B multiplicities within
a clan. The above formula simplifies considerably in the case of the
NBD, and even further if we assume that particles within a clan fall 
in F or B independently, since then $\gamma=(D_c^2-\nc)pq$:
\begin{equation}
	b =  \frac{2\beta p q }{1-2\beta p q }
\end{equation}
where $\beta$ is related to the average number of particles per clan:
$ \nc = \beta\left[(\beta-1)\ln(1-\beta)\right]^{-1} $.

We can now use experimental data to extract the values of the leakage
parameter $p$ which we do not know, then we use the NBD structure to
reproduce the almost linear relation between $\nbar_F$ and $n_B$.
Since at 63 GeV the mini-jet sample can be considered negligible, we
use only one component and obtain $p_{\text{soft}} = 0.78$. Because
the number of particles per clans in the soft component varies very
little from 63 to 900 GeV, we assume now that also the leakage
parameter is essentially constant in the GeV region. Using this value
for $p_{\text{soft}}$ and the measured value of the correlation
strength at 900 GeV, we can now calculate $p_{\text{semi-hard}} = 0.77$.
The just mentioned values can then be used
to predict the behaviour of the correlation strength with energy:
e.g., it can be shown that it will not continue to grow linearly 
with $\ln\sqrt{s}$ in the multi-TeV region
if the leakage parameter $p_{\text{semi-hard}}$ remains 
constant \cite{RU:FB}.

Within the same two-step process framework, we can extend our formulae
to the case in which either the reaction is asymmetric (e.g.\ consider
proton-nucleus or nucleus A-nucleus B with A $\ne$ B) or the F and B
regions are asymmetrically defined, or both cases are realised
\cite{RU:FBasymm}.
In order to proceed, we must allow for the possibility that 
the average leakage
from F to B may be different from the average leakage from B to F, and thus
introduce different parameters:
	$p_F \ne p_B$;
we must also allow for the possibility that, on average, more clans 
are produced in one region than in the other; we
call $r$ the average fraction of clans produced in the F region:
in general
	$r \ne 1/2$.
The symmetric case is recovered for $p_F = p_B$ and $r = 1/2$.

Using the formalism of generating functions, we arrive at formulae for
the joint MD in the general case. The joint generating function (GF)
for each component in Eq.~(\ref{eq:combo}) is
\begin{equation}
	G(z_F,z_B) = \sum_{n_F,n_B} z_F^{n_F} z_B^{n_B} P(n_F,n_B) .
\end{equation}
Since clans are Poisson distributed and independent of each other, the
joint GF for $N_F$ forward clans and $N_B$ backward clans is
\begin{equation}
	{\cal G}(z_F,z_B) = \exp\left\{ \Nbar [r z_F + (1-r) z_B - 1] \right\},
\end{equation}
where $\Nbar$ is the overall average number of clans.
Calling $g_{c,I}(z_F,z_B)$ the GF for the joint MD of particles from
one forward ($I=F$) or one backward ($I=B$) clan, we make use again of
the independence among clans and obtain the general formula:
\begin{equation}
	G(z_F,z_B) = 
		{\cal G} \left( g_{c,F}(z_F,z_B), g_{c,B}(z_F,z_B) \right).
																			\label{eq:14}
\end{equation}

The special case in which the MD is of NB type is of particular
interest.
In that case in fact one has simply (we assume that particles 
within a clan fall in F or B independently)
 \begin{equation}
	g_{c,F}(z_F,z_B) = \glog(z_F p_F  +  z_B q_F)
\qquad\text{and}\qquad
	g_{c,B}(z_F,z_B) = \glog(z_F q_B  +  z_B p_B) ,
\end{equation}
where $\glog$ is the GF for the logarithmic distribution of parameter
$\beta$:
\begin{equation}
	 \glog(z) \equiv \log(1-z \beta)/\log(1-\beta) .
\end{equation}
The generating function for 
the joint probability of $n_F$ and $n_B$ particles is then given by
\begin{equation}
	G(z_F,z_B) = \exp \left\{ 
				r\Nbar [ \glog(z_F p_F + z_B q_F) - 1 ]
			\right\} \exp \left\{
				(1-r)\Nbar [ \glog(z_F q_B + z_B p_B) - 1 ]
			\right\}   .
\end{equation}
This formalism can be used to describe FB  multiplicity correlations
in very general terms.
It is interesting to point out a few properties:
\begin{itemize}
\item The marginal distribution is the convolution of two NBD's, thus
it can be shown that it is an infinitely divisible distribution.
\item When $r=1/2$, one obtain for the GF of the marginal distribution a simple
correction to the NBD generating function; e.g., for the forward
marginal GF:
\begin{equation}
	g_F(z) \equiv G(z,1) =\left[ 1 + \frac{\nbar}{k}  (1-z)  + 
								\frac{\nbar^2}{k ^2}  p_F q_B (1- z) 
				\right]^{-k/2} .
\end{equation}

\item $\nbar_F(n_B)$ is not a linear function of $n_B$, and
vice-versa, except for the particular cases $p_i=1/2$ and $p_i=1$,
($i$=F,B); examples are shown in \cite{RU:FBasymm}.
\end{itemize}

In conclusion, general formulae for the 
FB multiplicity correlation strength and F and B joint distributions
generating functions have been given in the
framework of the superposition mechanism  of two weighted MD's
for different classes of events.
Assuming NB  regularity behaviour for 2- and 3-jet samples of events
in \ee\ annihilation at LEP energy, results obtained by OPAL
collaborations are correctly reproduced within experimental errors.
In  $pp$ collisions the FB multiplicity correlation
strengths in the two substructures (soft and semi-hard events) turn
out to be quite important:
in particular an interesting connection is found  between the
particle populations within clans, particle leakage from clans
in one hemisphere to the opposite hemisphere and superposition effect
between different substructure of the collision. This finding
favours structures with larger particle populations per clans and
the decrease of the average number of clans.  

It has been shown that assuming different particle leakage percentages
($p_B \ne p_F$) for binomially generated particles from clans in one
hemisphere to the opposite one and asymmetric ($r\ne 1/2$) distribution
in the two hemispheres of binomially generated clans, a general
formula for the generating function of the joint 
$(n_F,n_B)$-charged particle
multiplicity distribution for each class of events (or
substructure) in asymmetric reactions can be obtained when the total 
multiplicity distribution is of negative binomial type,
and can be easily generalised to any discrete
infinitely divisible multiplicity distribution.
This search is relevant for the study of forward-backward
multiplicity correlations in non-identical heavy ion 
collisions and in proton-nucleus collisions.
Accordingly, the newly introduced particle leakage and asymmetry
parameters can be used for
classifying different classes of collisions.

\bibliographystyle{prstyR}
\bibliography{abbrevs,bibliography}

\end{document}